\documentclass[twocolumn,english,prl,aps,amssymb,superscriptaddress,showpacs]{revtex4-1}
\usepackage[T1]{fontenc}
\usepackage{amsmath}
\usepackage{amssymb}
\usepackage{graphicx}

\makeatletter
\@ifundefined{textcolor}{}
{%
 \definecolor{BLACK}{gray}{0}
 \definecolor{WHITE}{gray}{1}
 \definecolor{RED}{rgb}{1,0,0}
 \definecolor{GREEN}{rgb}{0,1,0}
 \definecolor{BLUE}{rgb}{0,0,1}
 \definecolor{CYAN}{cmyk}{1,0,0,0}
 \definecolor{MAGENTA}{cmyk}{0,1,0,0}
 \definecolor{YELLOW}{cmyk}{0,0,1,0}
}

\usepackage{epsfig}
\usepackage{dcolumn}
\usepackage{bm}
\usepackage{babel}
\def\nn{\nonumber}

\begin{document}

\title{ Hidden Quasiparticles and Incoherent Photoemission Spectra in Na$_2$IrO$_3$ }

\author{Fabien Trousselet}
\affiliation{Max-Planck-Institut f\"ur Festk\"orperforschung,
             Heisenbergstrasse 1, D-70569 Stuttgart, Germany}
\affiliation{Institute N\'eel, CNRS/UJF, 25 Avenue des Martyrs, BP166,
             F-38042 Grenoble Cedex 9, France}

\author{Mona Berciu}
\affiliation{Department of Physics and Astronomy, University of British Columbia,
             Vancouver, British Columbia, Canada V6T 1Z1}
\affiliation{Quantum Matter Institute, University of British Columbia,
             Vancouver, British Columbia, Canada V6T 1Z4}

\author{Andrzej M. Ole\'s}
\affiliation{Max-Planck-Institut f\"ur Festk\"orperforschung,
             Heisenbergstrasse 1, D-70569 Stuttgart, Germany}
\affiliation{Marian Smoluchowski Institute of Physics, Jagellonian University,
             Reymonta 4, PL-30059 Krak\'ow, Poland}

\author{Peter Horsch}
\affiliation{Max-Planck-Institut f\"ur Festk\"orperforschung,
             Heisenbergstrasse 1, D-70569 Stuttgart, Germany}

\date{\today}

\begin{abstract}
We study two Heisenberg-Kitaev $t$-$J$-like models on a honeycomb
lattice, focusing on the zigzag magnetic phase of Na$_2$IrO$_3$,
and investigate hole motion by exact diagonalization and variational
methods. The spectral functions are quite distinct from those of
cuprates and are dominated by large incoherent spectral weight at
high energy, almost independent of the microscopic parameters ---
a universal and generic feature for zigzag magnetic correlations.
We explain why quasiparticles at low energy are strongly suppressed
in the photoemission spectra and determine an analog of a pseudogap.
We point out that the qualitative features of the predominantly
incoherent spectra obtained within the two different models for
the zigzag phase are similar, and they have remarkable similarity
to recently reported angular resolved photoemission spectra for
Na$_2$IrO$_3$.
\end{abstract}


\maketitle

Frustrated spin systems have long served as a rich source of exotic
phenomena such as unconventional magnetic order or spin liquid
behavior \cite{Bal10,Nor09}. A beautiful realization of a spin liquid
is found in the Kitaev model \cite{Kit06}, where bond-dependent Ising
interactions lead to strong frustration on the two-dimensional (2D)
honeycomb lattice. Here antiferromagnetic (AF) or ferromagnetic (FM)
couplings are equivalent; a FM realization was originally suggested
in Ref. \cite{Kha05}. The Kitaev model is exactly solvable and has
spin correlations that vanish beyond nearest neighbor (NN) spins
\cite{Bas07}. This spin liquid is robust against weak time-reversal
invariant perturbations, including the Heisenberg ones \cite{Cha10},
similar to the spin-orbital liquid in the SU(4) symmetric
Kugel-Khomskii model \cite{notekk}, but in striking contrast to
the 2D compass model \cite{Tro10}.

Such peculiar, highly  anisotropic interactions are  believed to be
realized in $A_2$IrO$_3$ ($A=$Na,Li) iridates where strong spin-orbit
coupling generates Kramers doublets of isospin states
\cite{Jac09,Shi09}. These systems are Mott insulators as confirmed by
the electronic structure \cite{Comin,notemo}.
It has been suggested that effective
$S=\frac12$ pseudospins, standing for locally spin-orbital-entangled
$t_{2g}$ states \cite{Hor03,Ole12}, interact via competing AF
Heisenberg and FM Kitaev exchange couplings between NNs in the
Heisenberg-Kitaev (HK) model --- this scenario is consistent with the
magnetic susceptibility \cite{Sin10} and with resonant inelastic x-ray
scattering (RIXS) \cite{Gre12}. Surprisingly, the theoretical
predictions of a spin liquid or stripy AF phase \cite{Cha10} were not
confirmed but instead zigzag AF order, consisting of staggered FM
zigzag chains, was found in Na$_2$IrO$_3$ \cite{Liu11,Sin12,Cho12,Ye12}.
To stabilize this phase next-nearest neighbor (NNN) and third nearest
neighbor (3NN) AF Heisenberg terms \cite{Alb11} have been invoked in
the HK model \cite{Kim11}. Recently a simpler scenario, including $e_g$
orbitals in the exchange paths, has been proposed \cite{You12,Cha12};
there, the zigzag AF phase emerges in a broad range of parameters in
the HK model when the signs of all NN exchange terms are {\it inverted}.

\begin{figure}[b!]
\includegraphics[width=3.5cm]{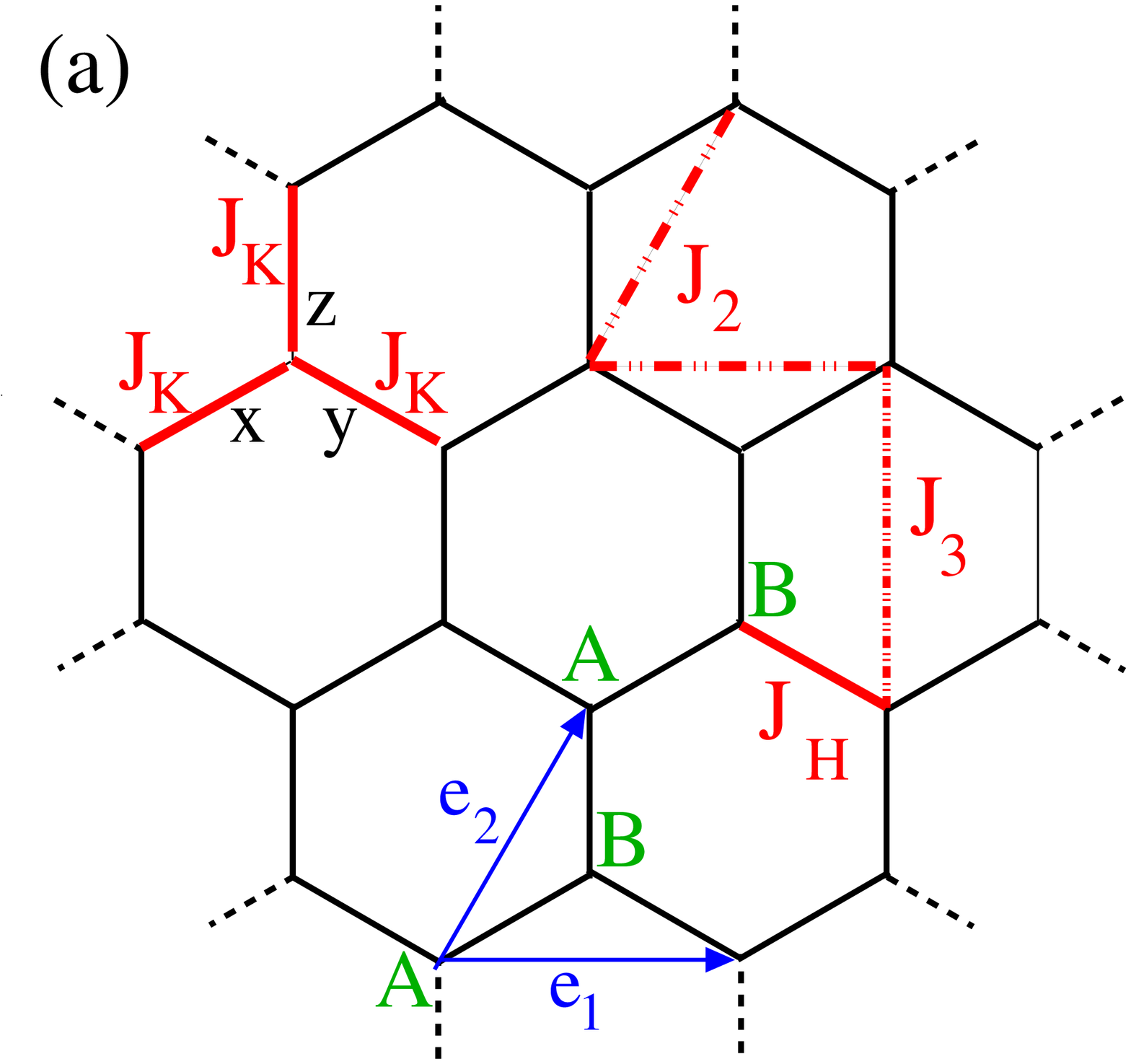}
\hskip .5cm
\includegraphics[width=3.7cm]{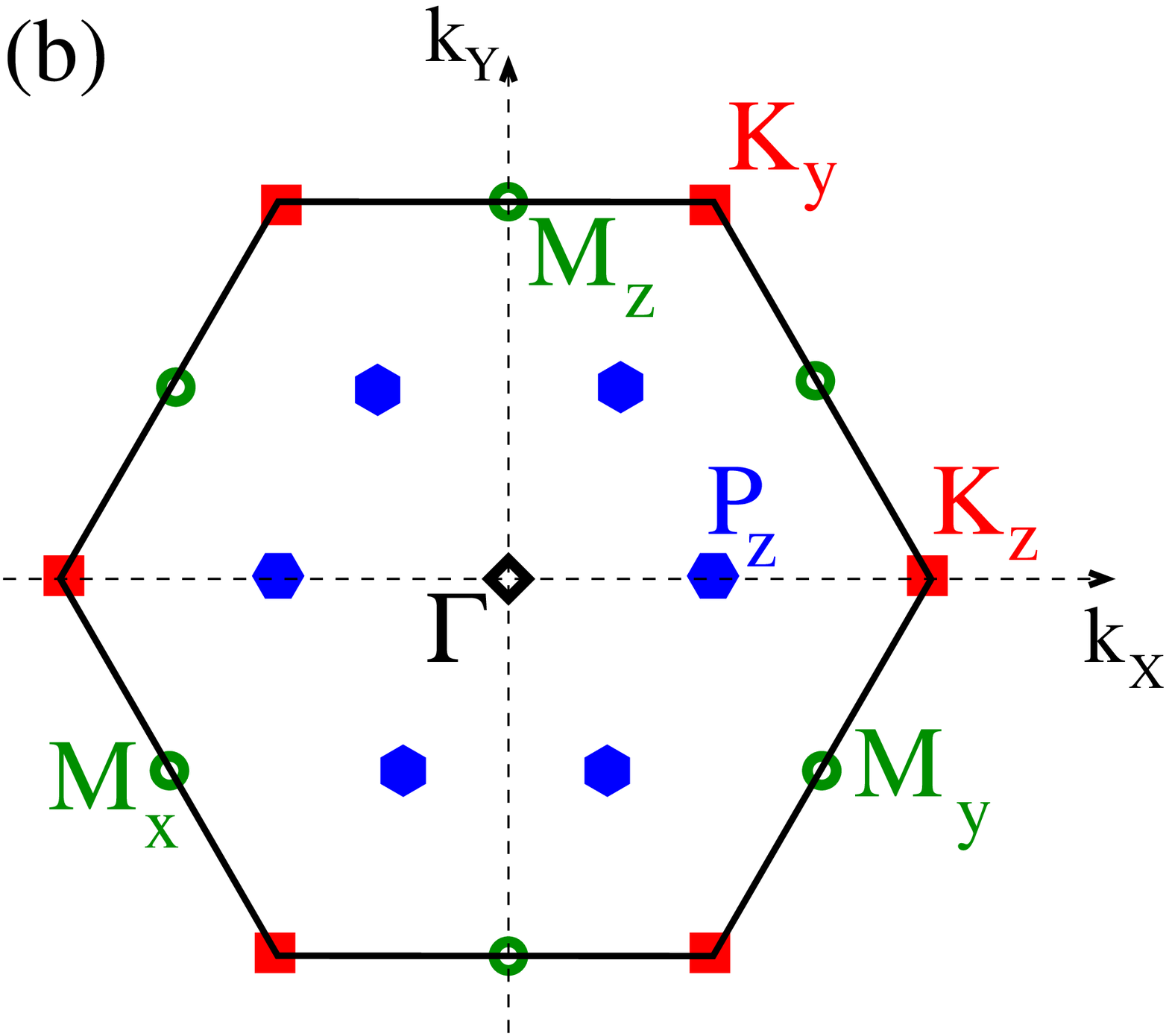}\\
\caption{(color online).
Honeycomb lattice of Na$_2$IrO$_3$: (a) cluster of $N=24$ sites,
and the elementary translations $\vec{e}_{1(2)}$
within two sublattices $A$ or $B$ (for two NN $A$ atoms $a=1$).
Exchange couplings are: in Heisenberg ($J_H$) and Kitaev ($J_K$)
(solid); Heisenberg NNN ($J_2$) and 3NN ($J_3$) (dashed).
(b)~First Brillouin zone with high symmetry points: $\Gamma$,
$M_y=(\pi,-\pi/\sqrt{3})$, $K_z=(4\pi/3,0)$; in exact diagonalization
$M_\gamma$ and $K_\gamma$ are equivalent ($\gamma\in\{x,y,z\}$) and
called $M$ and $K$.
}
\label{fig:defs}
\end{figure}

In this Letter we investigate whether hole motion in an effective HK
model can explain recent angle-resolved photoemission spectroscopy
(ARPES) experiments for Na$_2$IrO$_3$ \cite{Comin}.
One expects that a hole might move coherently along the FM zigzag
chain [Fig. 1(a)], similar to free hole propagation in states with
ferro-order of $t_{2g}$ orbitals \cite{Wro10}. It is therefore quite
surprising that the ARPES spectra for Na$_2$IrO$_3$ are dominated
instead by incoherent spectral weight found predominantly at high
energy \cite{Comin}. This poses several intriguing questions:
(i) Can one distinguish by ARPES the two zigzag states realized by
qualitatively different interactions?
(ii) Is there any similarity between the present ARPES results and
those known for cuprates \cite{Dam03}, described by quasiparticles
(QPs) within the $t$-$J$ model \cite{Mar91,Dag94}, or are the spectra
dominated by entangled spin-orbital excitations \cite{Woh09}?
(iii) Is the incoherent hole scattering on spin excitations as
important here as between the FM planes of LaMnO$_3$ \cite{Bal01}?
We answer them below and show that weak QPs exist but are hidden for
the present honeycomb lattice so that ARPES reveals the incoherent
processes.

We consider the following $t$-$J$-like model ($t>0$),
\begin{eqnarray}
{\cal H}_{tJ}&\equiv& {\cal H}_t+{\cal H}_J=
t\sum_{\langle ij\rangle\sigma} c^{\dagger}_{i\sigma}c^{}_{j\sigma} \nn\\
&+&J_H\sum_{\langle ij\rangle}   \vec{S}_i\cdot \vec{S}_j
 + J_K\sum_{\langle ij\rangle_\gamma} S_i^\gamma     S_j^\gamma\,.
\label{tJ}
\end{eqnarray}
on the honeycomb lattice [Fig. \ref{fig:defs}(a)], called {\it model I}.
The hopping ${\cal H}_t$ of composite fermions with pseudospin flavor
$\sigma$ \cite{Hya12} (contributions from direct $d$-$d$ and $d$-$p$-$d$
hopping \cite{Kha04}) occurs in the restricted space without double
occupancies due to large on-site Coulomb repulsion $U$, and the exchange
${\cal H}_J$ stands for the HK model ---
with FM Heisenberg ($J_H<0$) and AF Kitaev ($J_K>0$) exchange,
\begin{equation}
J_H\equiv -J(1-\alpha)\,, \hskip .7cm J_K\equiv 2J\alpha \,.
\label{JHK}
\end{equation}
Here $J$ is the energy unit and $\alpha\in [0,1]$ interpolates between
the Heisenberg and Kitaev exchange couplings for NN $\frac12$ spins
$\{\vec{S}_i\}$; the zigzag AF phase (Fig. 2) is found in a broad range
of $\alpha$ \cite{Cha12}. We also investigate the spectral functions in
{\it model II}, where exchange couplings $-{\cal H}_J$ are extended by
NNN and 3NN terms $J_n=J(1-\alpha)j_n>0$:
\begin{equation}
{\cal H}_{tJ}'= {\cal H}_t-{\cal H}_J+\Big\{
  J_2\!\!\!\! \sum_{\{ij\}\in{\rm NNN}}\!\!\!
  \vec{S}_i\!\cdot\!\vec{S}_j
+ J_3\!\!\!\! \sum_{\{ij\}\in{\rm 3NN}}\!\!\!
  \vec{S}_i\!\cdot\vec{S}_j\Big\}.
\label{j2j3}
\end{equation}

Although our aim is to present the spectral functions obtained by exact
diagonalization (ED) for a hole in the quantum-fluctuating zigzag AF
phase, we begin with describing the physical processes and resulting
spectral properties of a hole inserted into a fully-polarized ground
state $|0\rangle$, see Fig. 2(a). The hole hopping is
isotropic, but for convenience we distinguish {\it intrachain} hopping
$t$ and {\it interchain} hopping $t_{\perp}$. Free hole propagation,
see Fig. 2(b), occurs along the 1D FM zigzag chain when $t_{\perp}=0$.
It involves two types of Ir sites which belong to sublattices $A$
and $B$, see Fig. 1(a). We are interested in the spectral properties
measured in the ARPES experiment with the hole creation operator
$c^\dagger_{{\vec k}\uparrow}$, and we also consider the hole creation
on sublattice $A$, $d^{\dagger}_{{\vec k}\uparrow}$ \cite{Oleg}:
\begin{equation}
d_{{\vec k}\uparrow}^{\dagger}\!\equiv\sqrt{\frac{2}{N}} \sum_{i \in A}
e^{i \vec{k}\cdot\vec{r}_i} c_{i\uparrow}^{\dagger}, \hskip .3cm
c_{{\vec k}\uparrow}^{\dagger}\!\equiv\frac{1}{\sqrt{N}}\sum_i
e^{i\vec{k}\cdot\vec{r}_i} c_{i\uparrow}^{\dagger}.
\label{dc}
\end{equation}
As we shall see, the sublattice aspect is rather subtle and responsible
for the hidden QP states in the ARPES experiments for the zigzag phase.
The spectral functions
\begin{eqnarray}
\label{Af}
A_f(\vec k,\omega) &=& \frac{1}{\pi}\Im\langle 0|
f^\dagger_{{\vec k}\uparrow}\,\frac{1}{\omega-i\eta+E_0-{\cal H}}\,
f^{}_{{\vec k}\uparrow}|0\rangle ,
\end{eqnarray}
correspond to the physical Green's function $G_c({\vec k},\omega)$ for
$f\equiv c$, or to the sublattice Green's function
$G_d({\vec k},\omega)$ for $f\equiv d$ (\ref{dc}); here $E_0$~is the
energy of the ground state $|0\rangle$.

\begin{figure}[t!]
\includegraphics[width=8.0cm]{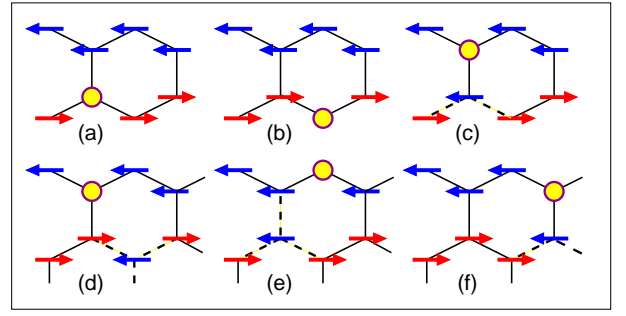}
\caption{(color online).
Artist's view of hole propagation in the zigzag phase of
Na$_2$IrO$_3$ (arrows). (a) The hole (circle) may either
(b) propagate along the FM chain by {\it intrachain} hopping $t$, or
(c) create a magnon by {\it interchain} hopping $t_{\perp}$. Either
this spin defect (d) or the hole (e) can move along its chain.
(f) After two steps (d)+(e) the hole and the defect may recombine.
Broken bonds are indicated by dashed lines.
}
\label{fig:art}
\end{figure}

Below we describe and use a variational approach well adapted to the
perturbative regime ($|t|\ll J$) to gain more physical insights.
We split the HK $t$-$J$ model Eq. (1) into an exactly solvable part
${\cal H}_0$ and the perturbation ${\cal V}$,
${\cal H}_{tJ}\equiv{\cal H}_0+{\cal V}$. When exchange interactions
are Isinglike, either in spin \cite{Tru88} or in orbital \cite{Wro10}
systems, the number of magnons (orbitons) is conserved and the
classical ground state $|0\rangle$ is exact.
Here we use the same strategy to develop a variational treatment
\cite{Ber11} and include in ${\cal H}_0$ all terms that conserve the
number of magnons, while ${\cal V}$ includes  the interchain hole
hopping which creates or removes a spin excitation [Fig. 2(c)] together
with exchange terms generating magnon pairs.

When a hole moves to a neighboring chain,
it creates a spin defect with energy $\varepsilon^{(0)}_m=|J_H|$, see
Fig. 2(c). The defect propagates as a magnon [Fig. 2(d)], and the hole
can also move by hopping $t$ [Fig. 2(e)] along the FM chain. Both
processes generate two parallel spins on a vertical (interchain) bond
and increase the magnon energy to
$\varepsilon_m=|J_H|+\frac12(J_K+J_H)$, but a hole and a spin defect
may also meet at one  vertical bond [Fig. 2(f)], which decreases the
magnon energy back to $\varepsilon^{(0)}_m$. All these processes cause
incoherence. Other states with several spin defects cost too much
energy when $J\gg t$ ---
these states are neglected in the one-magnon variational treatment.

\begin{figure}[t!]
\includegraphics[width=7.7cm]{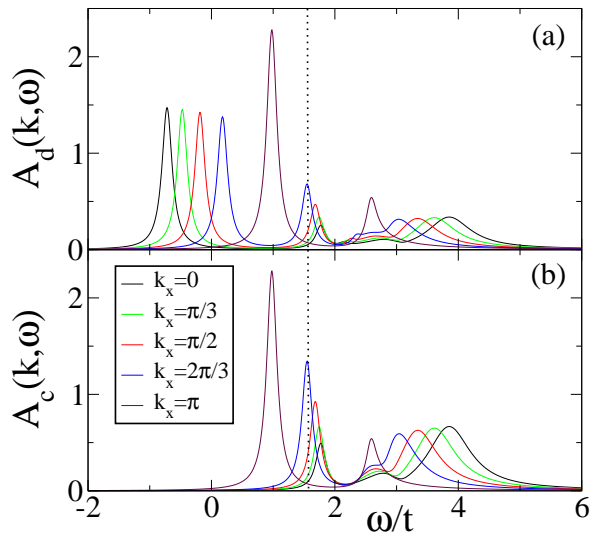}
\caption{(color online).
Spectral functions as obtained for model I (\ref{tJ}) in the
one-magnon approximation:
(a) $A_d(k_x,\omega)$ and
(b) $A_c(k_x,\omega)$ (\ref{Af}).
The dashed line indicates the static hole energy $\epsilon_0$.
Parameters: $\alpha=\frac{5}{9}$, $t/J=0.25$ and $\eta=0.1t$.
}
\label{fig:adac1}
\end{figure}

In the zigzag phase we write the Dyson's equation,
$G(\omega)=G^0(\omega)+G(\omega){\cal V}G^0(\omega)$,
for the resolvent,
$G(\omega)\equiv\{\omega-i\eta+E_0-{\cal H}\}^{-1}$.
The Green's functions $G_d({\bf k},\omega)$  obtained
from it are $2\times 2$ matrices for two sublattices, $A$ and $B$.
The unperturbed Green's function, $G_d^0({\bf k},\omega)$, is found
exactly --- it describes free propagation of a hole along the FM
chains, and depends on 1D momentum $k_x$ [Fig. 1(b)],
$\varepsilon_{{\vec k}\pm}=\epsilon_0\mp 2t\cos\frac12 k_x$, with
$\epsilon_0=\frac14 J(1+\alpha)$ being the energy of a static hole.
The two $\pm$ states for each $k_x$ result from band folding
associated with the two-site unit cell, independent of the 1D nature
of hole motion. These states give two QPs at each momentum $k_x$ in
$A_d^{(0)}({\bf k},\omega)$, except at $k_x=\pi$ where
$\varepsilon_{{\vec k}\pm}=\epsilon_0$.
In contrast, the lower
energy QP, $\varepsilon_{{\vec k}+}$, is {\it hidden} in
$A_c^{(0)}({\bf k},\omega)$, while the higher energy one,
$\varepsilon_{{\vec k}-}$, has twice larger spectral weight,
due to interference between the two sublattice contributions.

The difference between
$A_d({\vec k},\omega)$ and $A_c({\vec k},\omega)$ (\ref{Af})
for model I which follows from states parity is well visible
in the weak coupling regime of $J\gg t$, see Fig.~\ref{fig:adac1}
\cite{note59}. At low energy, $\omega<\epsilon_0$, one finds QPs
in $A_d(k_x,\omega)$, but not in $A_c(k_x,\omega)$.
The width of the QP band is somewhat reduced from the expected $2t$,
but even stronger renormalization (its width is lower than $t$)
is found for $\omega>\epsilon_0$.
New incoherent features observed in both spectral functions
(\ref{Af}) above $\omega=2t$ (Fig.~\ref{fig:adac1}) are generated by
a magnon excitation due to the interchain hopping $t_{\perp}$, see
Fig. 2(c). These states disperse on the scale of $\sim 2t$
(with energy difference between the respective maxima at $k_x=0$ and
$\pi$ being $\simeq 1.4t$) and we recognize here the propagation of
a {\it holon} \cite{Dag07,Phi10}.

\begin{figure}[t!]
\includegraphics[width=7.7cm]{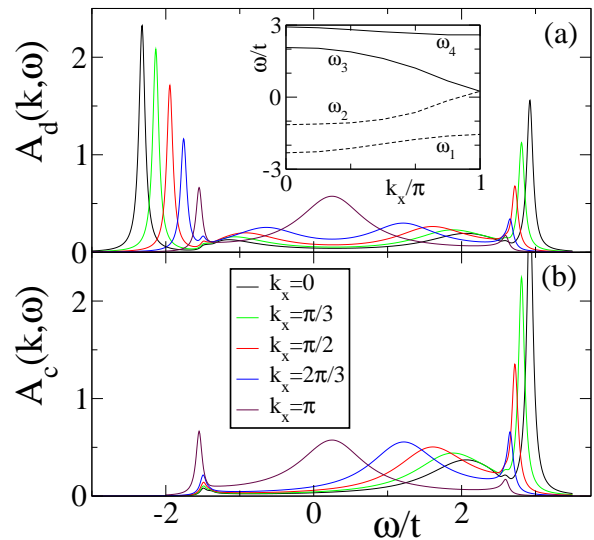}
\caption{(color online).
Spectral functions as obtained for model
I (\ref{tJ}) in the one-magnon approximation:
(a) $A_d(k_x,\omega)$, and
(b) $A_c(k_x,\omega)$ (\ref{Af}).
The inset shows $k_x$-dependence of the frequencies $\omega_i$
corresponding to the four distinct features visible in
$A_d(k_x,\omega)$;
the $\omega_1$ and $\omega_2$ ones are hidden in $A_c(k_x,\omega)$.
Parameters: $\alpha=\frac{5}{9}$, $t/J=2.0$ and $\eta=0.1t$.
}
\label{fig:adac2}
\end{figure}

In the intermediate coupling regime $t>J$, the spectral weight moves to
higher energies (Fig. \ref{fig:adac2}). First, one recognizes distinct
QPs in $A_d({\vec k},\omega)$ [{\it hidden} in $A_c({\vec k},\omega)$],
with further decreased total width of this subband and the spectral
weight decreasing from $k_x=0$ to $\pi$. This follows from hole
scattering on magnon excitations (Fig. 2) that become gradually more
important when $t/J$ increases.
Second, the spectra $A_c({\vec k},\omega)$ at $t/J=2.0$ [Fig.
\ref{fig:adac2}(b)] have three notable features:
(i) the suppressed QP peaks at the onset of the spectrum,
(ii) the upper holon branch, with spectral weight moving to higher
energies when $k_x$ decreases from $\pi$ to $0$, and
(iii) large spectral weight with weak momentum dependence at the upper
edge of the spectrum.
These higher energy features are expected to be strongly renormalized
when the constraint to one-magnon excitations is lifted. In contrast,
the QPs disappear in the strong coupling regime $t\gg J$ as well,
because the destructive interference is due to parity.

\begin{figure}[t!]
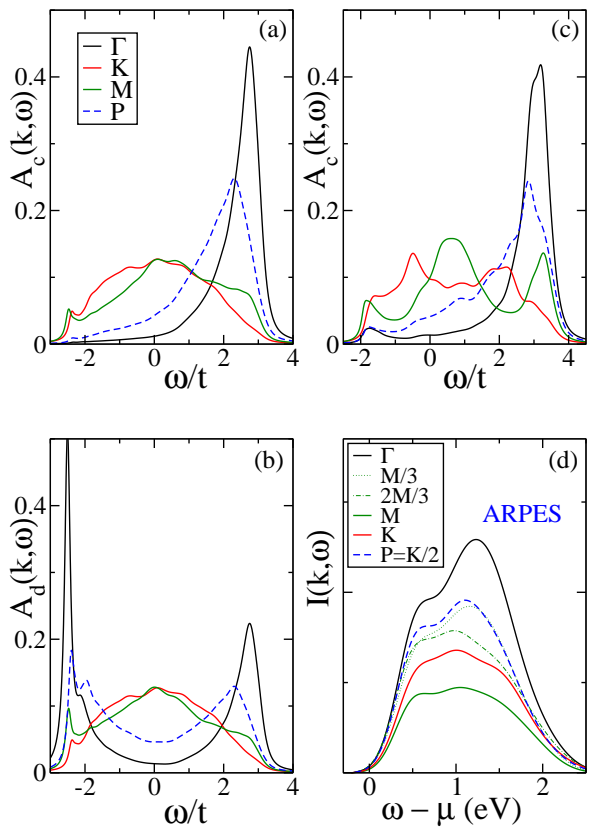

\includegraphics[width=7.7cm]{spec24ac}
\vskip .6cm
\includegraphics[width=7.7cm]{spec24bd}
\caption{(color online).
Spectral functions as obtained by ED for $N=24$ sites at selected
high-symmetry points $\Gamma$, $K$, $M$, and $P$ in the BZ at $t/J=5$
and $\eta=0.1t$:
(a) $A_c(k,\omega)$ and
(b) $A_d(k,\omega)$,
both for model I (1) with $\alpha=\frac{5}{9}$;
(c) $A_c(k,\omega)$ for model II (3), with $\alpha=0.4$, $j_2=0.2$ and
$j_3=0.5$.
ARPES spectra for Na$_2$IrO$_3$ \cite{Comin} with background subtracted
\cite{notesw} are shown in (d) for the selected ${\vec k}$ points
\cite{notear}.
}
\label{fig:ed}
\end{figure}

Indeed, the unbiased ED \cite{noteed} performed on a periodic cluster
of $N=24$ sites [Fig. 1(a)] for model I with the same
$\alpha=\frac{5}{9}$ yields very incoherent spectral weight
distribution in $A_c({\vec k},\omega)$, see Fig.~\ref{fig:ed}(a),
in contrast to the 2D $t$-$J$ model \cite{Dag94,vSz91}. As
most important feature, the spectral weight is moved to high energy
for the $P$ and $\Gamma$ points. At the $M$ point one observes a broad
feature at $\omega/t\simeq 0.3$, accompanied by a shoulder at
$\omega\simeq 2.5t$, and a small QP peak at $\omega\simeq -2.5t$ ---
such a peak is also observed at $K$ but absent for momenta inside the
Brillouin zone (BZ). The spectral weight is transferred from high to
lower energy when one moves from the $\Gamma$ point to the edge of the
BZ ($M$ and $K$ point). These qualitative features are generic and hold
in a broad range of parameters (see the Appendix).

The spectral weight is distributed quite differently in
$A_d({\vec k},\omega)$ [Fig. \ref{fig:ed}(b)]. Here QP peaks appear for
all momenta at energies $\omega <-2t$, with much more weight than those
observed in $A_c({\vec k},\omega)$, and there is strong spectral weight
transfer to low energies at $P$ and $\Gamma$ points. The QP weight is
maximal for the $\Gamma$ point and small for $K$ and $M$, as expected
from the variational approach [Fig. \ref{fig:adac2}(a)]. We suggest
that the lowest energy QP at the $\Gamma$ point determines the chemical
potential $\mu$ and is responsible for the experimentally seen
{\it pseudogap\/} of $\Delta\sim 0.35 eV$ \cite{Comin}.

It is quite remarkable that the ED results obtained with model II
[Fig.~\ref{fig:ed}(c)] for parameters suggested \cite{Cho12} for
Na$_2$IrO$_3$, are qualitatively similar to those found in model I
[Fig. \ref{fig:ed}(a)], but have somewhat richer structure. At the $M$
point one finds a two-peak structure with a weak QP at the low energy
side, which develops to a shoulder at the $K$ point. The spectra found
at $P$ and $\Gamma$ points look similar to those of model I, again with
most of the weight concentrated at high energy. The broad incoherent
spectral weight and its shift to higher energy between the $K$ and
$\Gamma$ point
are recognized here as universal features for the parameters favoring
zigzag phase. Indeed, moderate changes of parameters ($\alpha$, and
for model II $j_2$ and $j_3$) modifying the degree of frustration
while still supporting the zigzag order, do not lead to emerging QPs
(see the Appendix).

We suggest that the ED \cite{noteed} simulates here the main features
of the ARPES spectra measured at $T=130$ K \cite{Comin} --- although
the N\'eel temperature $T_N\simeq 15$~K is much smaller due to
frustration, the incoherent part of the spectra should be only weakly
affected by thermal fluctuations for $T \lesssim J$. The spectra are
incoherent and no QPs are seen at the low edge $\omega\simeq\mu$, see
Fig. \ref{fig:ed}(d).
With the total width of the spectrum $\sim 6t$ we estimate that
$t\simeq 0.3$ eV. Large spectral weight at the $\Gamma$ and $P$ points
is seen mostly at high energy, as in the ED in $A_c({\vec k},\omega)$
but not in $A_d({\vec k},\omega)$ [Figs. \ref{fig:ed}(a) and
\ref{fig:ed}(b)]. At $K$ and $M$ one recognizes three characteristic
features with appreciable spectral weight in the ARPES
data \cite{Comin} at $\omega-\mu\simeq 0.6$ \cite{notej},
$\simeq 1.1$, and $\simeq 1.5$ eV which have some correspondence in
the ED spectra. The basic features of the ARPES spectra,
(i)~the strong incoherence and
(ii)~the shift of the spectral weight to high energies at the $P$ and
$\Gamma$ points, are dictated by the zigzag correlations and are similar
in the two models. Yet, we notice differences in the fine-structure
which reflect the dramatically different underlying Hamiltonians,
and therefore suggest that model I is closer to the experimental data
\cite{Comin}.

Summarizing, we have shown that essential features seen in recent
ARPES experiments for Na$_2$IrO$_3$ \cite{Comin} may be described by a
universal phenomenology which does not depend on details of modeling.
The framework  used is a $t$-$J$-like model with nearest neighbor
Kitaev and Heisenberg exchange, having conceptually some similarity
to high $T_c$ cuprates. The physical spectral function
$A_c({\vec k},\omega)$ in the strong coupling regime explains
qualitatively the incoherent nature of the ARPES spectra, with large
spectral weight at high energy and the pseudogap determined by
{\it hidden\/} quasiparticles at low energy.
This also confirms that the low-energy electronic structure of
Na$_2$IrO$_3$ can be described by the motion of composite $j=\frac12$
fermions due to strong spin-orbit coupling.

{\it Acknowledgments.---}
We thank Andrea Damascelli, Lou-Fe' Feiner, George Jackeli,
Giniyat Khaliullin, George A. Sawatzky, and Ignacio Hamad for
insightful discussions. We are grateful to Riccardo Comin for
providing the experimental data of Ref.~\cite{Comin}.
This work was supported by
the Max Planck --- UBC Centre for Quantum Materials; F.T. thanks
Advanced Materials and Process Engineering Laboratory (AMPEL),
University of British Columbia, Vancouver for kind hospitality.
A.M.O. kindly acknowledges support by the Polish National
Science Center (NCN) under Project No. 2012/04/A/ST3/00331.

\section{APPENDIX: PARAMETER DEPENDENCE}

This Appendix presents additional numerical evidence
which demonstrates that the results of exact diagonalization (ED)
presented in Fig. 5 for the two models considered in the Letter
are robust. In each case the spectra only weakly depend on the
model parameters in the regime of the zigzag magnetic phase.

For convenience, we repeat first the definitions of the models
considered in the ED calculations. We consider
the following $t$-$J$-like {\it model I},
\begin{eqnarray}
{\cal H}_{tJ}&\equiv& {\cal H}_t+{\cal H}_J=
t\sum_{\langle ij\rangle\sigma} c^{\dagger}_{i\sigma}c^{}_{j\sigma}
\nonumber\\
&+&J_H\sum_{\langle ij\rangle}   \vec{S}_i\cdot \vec{S}_j
 + J_K\sum_{\langle ij\rangle_\gamma} S_i^\gamma     S_j^\gamma\,.
\label{tJmodel}
\end{eqnarray}
on the honeycomb lattice.
The operators $c^{\dagger}_{i\sigma}$ in the hopping term ${\cal H}_t$,
with $t>0$, stand for creation of composite fermions with pseudospin
flavor $\sigma$ in the restricted space without double occupancies;
the constraint follows from large on-site Coulomb repulsion $U$.
The exchange term ${\cal H}_J$ describes in this case the
Heisenberg-Kitaev model, with ferromagnetic (FM) Heisenberg ($J_H<0$)
and antiferromagnetic (AF) Kitaev ($J_K>0$) exchange, given by the
interaction $J$ and $\alpha\in [0,1]$, see Eqs. (\ref{JHK}).
The parameter $\alpha$ interpolates between the Heisenberg
($\alpha=0$) and Kitaev ($\alpha=1$) exchange couplings acting between
pairs of nearest neighbor (NN) spins $S=\frac12$; zigzag AF order is
found in a broad range of $\alpha$ \cite{Cha12}.

The second model investigated in the Letter, called {\it model II}, has
inverted signs of the NN spin couplings in the $-{\cal H}_J$ term,
and contains in addition exchange interactions between next NN (NNN)
and third NN (3NN) spins, $J_n=J(1-\alpha)j_n>0$ with $n=2,3$
\cite{Kim11}:
\begin{equation}
{\cal H}_{tJ}'= {\cal H}_t-{\cal H}_J+\Big\{
  J_2\!\!\!\! \sum_{\{ij\}\in{\rm NNN}}\!\!\!
  \vec{S}_i\!\cdot\!\vec{S}_j
+ J_3\!\!\!\! \sum_{\{ij\}\in{\rm 3NN}}\!\!\!
  \vec{S}_i\!\cdot\vec{S}_j\Big\}.
\label{J2J3}
\end{equation}
Further neighbor Heisenberg interactions $\{J_2,J_3\}$ are necessary
to stabilize zigzag AF order which is unstable in case of only NN AF
Heisenberg and FM Kitaev interactions (at $J_2=J_3=0$) \cite{Cha12}.

\begin{figure}[t!]
\begin{centering}
\begin{center}
\includegraphics[width=8cm]{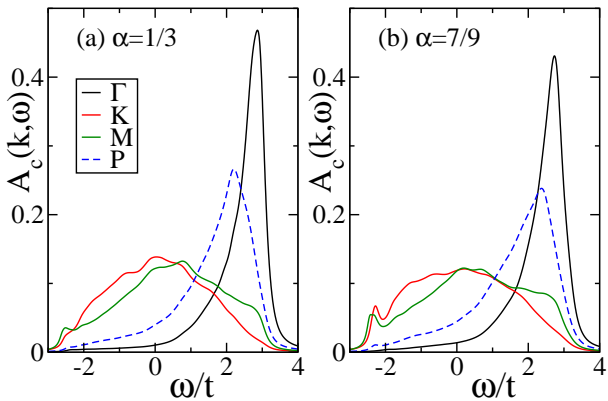}
\end{center}
\par\end{centering}
\caption{
Spectral functions $A_c(k,\omega)$ as obtained by ED for a periodic
cluster of $N=24$ sites at selected high-symmetry points $\Gamma$,
$K$, $M$, and $P$ in the BZ at $t/J=5$ and $\eta=0.1t$ for model I
Eq. (1), with:
(a) $\alpha=\frac{1}{3}$, and
(b) $\alpha=\frac{5}{9}$.
}
\label{fig:1}
\end{figure}

The ED calculations were performed on a periodic cluster of $N=24$
sites for model I and model II for the parameters favoring the expected
zigzag magnetic order in the ground state. Note that spin correlations
are systematically reduced, compared to those in a zigzag-ordered
phase, by the fact that no symmetry is spontaneously broken in the
ground state of a finite cluster. First, we vary the
parameter $\alpha$ and change the balance between the FM Heisenberg
and AF Kitaev interactions in Eqs. (\ref{JHK}) from $\alpha=\frac13$
to $\alpha=\frac{7}{9}$. As the energy increases in this range with
increasing $\alpha$ \cite{Cha12}, frustration of spin interactions
in Eq. (\ref{tJ}) increases as well. Surprisingly, the spectra
presented in Fig. 6 and in Fig. 5(a) for the selected
${\vec k}$ points in the Brillouin zone (BZ) are only rather
weakly influenced by that. The overall width of somewhat less than
$6t$ does not change showing that it is determined by the hopping
term ${\cal H}_t$. The spectral
weight is mostly incoherent --- it concentrates in the central
intermediate energy part for the $M$ and $K$ points at the edge of the
BZ, while it is moved to high energy for the $P$ and $\Gamma$ points.
A weak low energy quasiparticle (QP) is visible at the $M$ and $K$
points with the spectral weight growing with $\alpha$. But even at
large $\alpha=\frac{7}{9}$ these QPs are rather weak and incoherent
spectral weight dominates.
The two cases displayed in Fig. 6 are nearly identical to each other
and to the representative spectrum for $\alpha=\frac{5}{9}$ presented
in Fig. 5(a).

\begin{figure}[t!]
\begin{centering}
\begin{center}
\includegraphics[width=8cm]{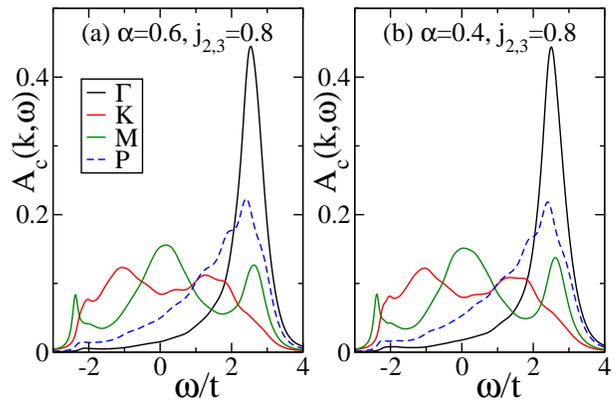}
\end{center}
\par\end{centering}
\caption{
Spectral functions $A_c(k,\omega)$ as obtained by ED for $N=24$ sites
at selected high-symmetry points $\Gamma$, $K$, $M$, and $P$ in the BZ
at $t/J=5$ and $\eta=0.1t$ for model II Eq. (3), with $j_2=j_3=0.8$,
and:
(a) $\alpha=0.6$,
(b) $\alpha=0.4$.
}
\label{fig:2}
\end{figure}

Second, we consider model II Eq. (\ref{J2J3}) and remark that the ED
spectra obtained with it are qualitatively similar to those of model I
at the same points in the BZ, see Fig. 7.
Again, the spectral weight is mostly incoherent and moved to high
energy at the $\Gamma$ and $P$ points. Here we consider two
interaction parameter sets $\{\alpha,j_2,j_3\}$ which have less
frustration than the case of $\alpha=0.4$, $j_2=0.2$ and $j_3=0,5$
shown in Fig. 5(c), and the zigzag magnetic order is
even more favored. While the spectral functions at the $\Gamma$
and $P$ points are nearly insensitive to these parameter changes,
a weak QP forms and is better resolved at the $M$ point for both
parameter sets of Fig. 7 than at $\{\alpha,j_2,j_3\}=\{0.4,0.2,0.5\}$.
Again, the spectra found at the points belonging to the edge of the BZ
($K$ and $M$) have richer structure than those in model I (Fig. 6).
At the $M$ point a two-peak structure is found, with a weak QP forming
a shoulder of the lower peak at the low energy side. The broad
incoherent spectral weight and its shift to higher energy between the
edge and the center of the BZ are recognized as universal features
for the parameters favoring zigzag magnetic order.

Summarizing, the numerical data show that for both models the
suppression of low energy QPs occurs in a broad range of parameters
which correspond to the zigzag magnetic order in the ground state,
and is therefore not a result of parameter tuning. This makes the
observations made in the Letter robust features of the presented
models I and II. We have found:
(i) the incoherent distribution of the spectral weight,
(ii) large spectral weight at high energy, and
(iii) the absence of low energy QPs from the ARPES spectra because of
destructive interference, even though QP eigenstates exist in the
low-energy spectrum. This could explain pseudogap-type behavior in
these materials.

These results suggest that the qualitative agreement with the recently
reported experimental angle-resolved photoemission spectroscopy (ARPES)
spectra \cite{Comin} is generic, and thus bring new insights into
spectral properties of such a strong spin-orbit-coupled Mott insulator.


\begin{thebibliography}{99}
\bibitem{notej} The ED does not reproduce this low energy excitation
                with no ${\vec k}$ dependence which might correspond
                to $j=\frac{3}{2}$ states, not included in
                the effective $t$-$J$-like model Eq. (1).
\bibitem{notesw} The norm of $A_c({\vec k},\omega)$ in ED is $\frac12$
                at half-filling as expected for constrained fermions.
                Due to matrix elements and background subtraction
                $I({\vec k},\omega)$ obeys no sum rule.
\bibitem{notear} As in our definition of $A({\vec k},\omega)$ Eq.
                 (\ref{Af}), we plot increasing ARPES excitation
                 energies in $I({\vec k},\omega)$ from left to right.
\end{thebibliography}
\end{document}